\catcode`\@=11
\def\citen#1{\if@filesw \immediate\write \@auxout {\string\citation{#1}}\fi%
\@tempcntb\m@ne \let\@h@ld\relax \def\@citea{}%
\@for \@citeb:=#1\do {\@ifundefined {b@\@citeb}%
    {\@h@ld\@citea\@tempcntb\m@ne{\bf ?}%
    \@warning {Citation `\@citeb ' on page \thepage \space undefined}}%
    {\@tempcnta\@tempcntb \advance\@tempcnta\@ne
    \setbox\z@\hbox\bgroup\ifcat0\csname b@\@citeb \endcsname \relax
    \egroup \@tempcntb\number\csname b@\@citeb \endcsname \relax
    \else \egroup \@tempcntb\m@ne \fi \ifnum\@tempcnta=\@tempcntb
    \ifx\@h@ld\relax \edef \@h@ld{\@citea\csname b@\@citeb\endcsname}%
    \else \edef\@h@ld{\hbox{--}\penalty\@highpenalty
    \csname b@\@citeb\endcsname}\fi
    \else \@h@ld\@citea\csname b@\@citeb \endcsname \let\@h@ld\relax \fi}%
\def\@citea{,\penalty\@highpenalty\hskip.13em plus.13em minus.13em}}\@h@ld}
\def\@citex[#1]#2{\@cite{\citen{#2}}{#1}}%
\def\@cite#1#2{\leavevmode\unskip\ifnum\lastpenalty=\z@\penalty\@highpenalty\fi%
  \ [{\multiply\@highpenalty 3 #1%
  \if@tempswa,\penalty\@highpenalty\ #2\fi}]}   %
\makeatother 

\newcounter{fnot}

\def\one           {\psi}
\def\two           {{\psi^+}}
\def\three         {\varepsilon}
\def\four          {\sigma}
\def\five          {{\sigma^+}}
\def\vier          {[44]}
\def\vier          {\eta}
\def\zwei          {[22]}
\def\zwei          {\xi}

\def\alg           {algebra}
\def\auto          {automorphism}

\def\bc            {boundary condition}
\def\be            {\begin{equation}}
\def\bearl         {\begin{array}{l}}
\def\bearll        {\begin{array}{ll}}

\def\caD           {\mbox{$\wfont D$}}
\def\CAD           {{({\wfont D})}}

\def\caN           {\mbox{$\wfont N$}}
\def\cb            {chiral block}
\def\cc            {{\rm C}}

\def\cft           {conformal field theory}
\def\cfts          {conformal field theories}
\def\chii          {\raisebox{.15em}{$\chi$}}

\def\chir          {\mbox{$\wfont A$}}
\def\chira         {chiral algebra}
\def\chirw         {\mbox{$\cal W$}}

\def\dim           {dimension}
\def\dl            {\mathbb }

\def\ee            {\end{equation}}

\def\eear          {\end{array}}
\def\emt           {energy-momentum tensor}
\def\eq            {\,{=}\,}
\newcommand\erf[1] {(\ref{#1})}

\def\findim        {finite-dimensional}

\def\hil           {\mbox{$\cal H$}}
\def\hJ            {\mbox{${\cal H}_\J$}}
\def\hl            {\mbox{${\cal H}_\lambda$}}
\def\hL            {{\cal H}_\lambda}
\def\hlo           {\mbox{${\cal H}_{\om(\lambda)}$}}

\def\hvac          {\mbox{${\cal H}_\vac$}}
\def\hvaca         {\mbox{${\cal H}_{(11)}$}}
\def\hvacA         {\mbox{${\cal H}_\vacA$}}
\def\hwa           {\mbox{${\cal H}_{(41)}$}}
\def\hwb           {\mbox{${\cal H}_{(21)}$}}
\def\hwc           {\mbox{${\cal H}_{(31)}$}}
\def\hws           {highest weight sta\-te}
\def\hwv           {highest weight vec\-tor}
\def\hy            {$\mbox{-\hspace{-.66 mm}-}$}
\def\id            {{\rm id}}
\def\Id            {{\rm id}}
\def\ii            {{\rm i}}
\def\iN            {\,{\in}\,}
\def\irrep         {irreducible representation}
\def\J             {{\rm J}}
\def\ja            {{(41)}}
\long\def\labl#1   {\label{#1}\ee \ifnum\draftcontrol=1
                   \mbox{ }\\[-12 mm]\query{#1}\\[5 mm] \fi}
\def\lambdad       {{\dot\lambda}}

\def\liefont       {\mathfrak }

\newcommand\N[3]   {{\rm N}_{#1,#2}^{\;\ \ #3}}
\newcommand\NA[3]  {{}_{}^{\rm(A)}{\rm N}_{\dot #1,\dot #2}^{\;\ \ \dot #3}}
\newcommand\NAJ[3] {{}_{}^{\rm(A)}{\rm N}_{\dot #1,\dot #2}^{\;\;\ \J\star
                   \dot #3}}
\newcommand\ND[3]  {{}_{}^\CAD{\rm N}_{#1,#2}^{\;\ \ #3}}
\def\nE            {\,{\not=}\,}
\def\NE            {\mbox{$N_{{\rm A}\not\to{\rm D}}$}}
\newcommand\NI[3]  {{}_{}^\tci{\rm N}_{\dot #1,\dot #2}^{\;\ \ \dot #3}}
\newcommand\NJ[3]  {{}_{}^\tci{\rm N}_{\dot #1,\dot #2}^{\;\,\ (41)\star
                   \dot #3}}
\def\NO            {\mbox{$N_{{\rm A\to D,red.}}$}}
\newcommand\Ntot[3]{{\rm M}^{#1,#2}_{\;\ \ #3}}
\newcommand\nxl[1] {\\{}\\[-.#1em]}
\newcommand\Nxl[2] {\\{}\\[-#1.#2em]}

\def\om            {\omega}
\def\Om            {\omega^\circ}
\def\oml           {\omega_\lambda}
\def\Oml           {\Om_\lambda}
\def\omvac         {\omega_\vac}
\def\onedim        {one-dimen\-sional}

\def\P             {{\phantom|}}
\def\pJ            {\phi_{\rm J}}
\def\pja           {\phi_{(41)}}

\newcommand\rc[3]  {R^{#1}_{#2,\om(#2);#3}}

\newcommand\rcA[2] {R^{(\pi{\circ}\Om;#1)}_{#2,#2;\vac}}
\newcommand\rcc[3] {R^{(\cc;#1)}_{#2,#2^+_{\phantom|};#3}}
\newcommand\rci[2] {R^{(\Id;#1)}_{#2,#2;\vac}}
\def\rep           {representation}
\def\rhs           {right hand side}
\def\q             {quantum }

\def\resp          {respectively}
\newcommand\SA[2]  {S^{\rm(A)}_{\dot #1,\dot #2}}

\newcommand\SAV[1] {S^{\rm(A)}_{\dot #1,\dot\vac}}

\newcommand\SD[2]  {S_{#1,#2}}

\newcommand\SI[2]  {S^\tci_{\dot #1,\dot #2}}
\newcommand\SIv[1] {S^\tci_{\dot #1,\vaca}}

\def\smat          {S-matrix}

\def\tci           {{\scriptscriptstyle\rm(Is\,4)}}
\def\teim          {tetracritical Ising model}

\def\tsp           {three-sta\-te Potts mo\-del}
\def\twodim        {two-dimen\-sio\-nal}
\def\vac           {\Omega}
\def\vaca          {{(11)}}
\def\vacA          {{\vac^{\rm(A)}}}
\newcommand\version[1] {\ifnum\draftcontrol=1 \typeout{}\typeout{#1}\typeout{}
                   \vskip3mm \centerline{\fbox{{\tt DRAFT -- #1 -- }
                   {\small\draftdate}}}
                   \vskip3mm \fi}
\def\vir           {\mbox{$\liefont{Vir}$}}
\def\vira          {Virasoro algebra}
\def\wfont         {\mathfrak }
\def\wrt           {with respect to }
\def\wrtt          {with respect to the }

\def\wzwt          {WZW theory}
\def\wzwts         {WZW theories}
\def\zet           {{\dl Z}}

\global\def\draftcontrol{0}
\catcode`\@=12

     \documentclass[12pt]{article} 
     \usepackage{amssymb,amsfonts}
\begin{document}


  \begin{flushright}
  {~} \\[-15 mm]  {\sf hep-th/9806121} \\[1mm]
  {\sf CERN-TH/98-163} \\[1 mm]
  {\sf June 1998} 
  \end{flushright}
  \begin{center} \vskip 15mm
  {\Large\bf COMPLETENESS OF BOUNDARY CONDITIONS}\\[4mm]
  {\Large\bf FOR THE CRITICAL THREE-STATE POTTS MODEL}\\[16mm]
  {\large J\"urgen Fuchs} \\[3mm]
  Max-Planck-Institut f\"ur Mathematik \\[.6mm]
  Gottfried-Claren-Str.\ 26, \  D -- 53225~~Bonn \\[11mm]
  {\large Christoph Schweigert} \\[3mm] CERN \\[.6mm] CH -- 1211~~Gen\`eve 23
  \end{center}
  \vskip 20mm
  \begin{quote}{\bf Abstract}\\[1mm]
We show that the conformally invariant boundary conditions for 
the three-state Potts model are exhausted by the eight known solutions.
Their structure is seen to be similar to the one in a free field theory
that leads to the existence of D-branes in string theory.
Specifically, the fixed and mixed boundary conditions 
correspond to Neumann conditions, while the free boundary
condition and the new one recently found by Affleck et al.\ \cite{afos} 
have a natural interpretation as Dirichlet conditions for a higher-spin 
current. The latter two conditions are governed by the Lee\hy Yang fusion rules.
These results can be generalized to an infinite series of non-diagonal minimal 
models, and beyond.
  \end{quote}        
  \newpage
 

The simplest \bc s for a \cft\ are those
which preserve not only the conformal symmetry, but also all other
symmetries of the model. In other words, not only do they leave the Virasoro 
\alg\ \vir\ invariant, but even the whole chiral \alg\ \chir. 
As already argued long ago by Cardy \cite{card9}, for 
models in which the torus partition function is given by charge conjugation,
such \bc s are governed by the fusion rule \alg\ \caN\ of the model.
In particular, the possible \bc s are in one-to-one correspondence to the
primary fields (\wrt \chir).

When the chiral \alg\ \chir\ is larger than \vir, then
in general there also exist conformally invariant \bc s that
do {\em not\/} preserve all of \chir.
An obvious task is then to identify structures that govern the \bc s in
this more general situation. This problem can be attacked by identifying 
suitable compatibility requirements which enforce that the \bc s for
the fields in the extended \alg\ cannot be chosen arbitrarily. For instance,
for free bosons and for \wzwts\ such constraints arise from the fact that the 
\emt\ is a quadratic expression in the (spin-1) currents.
Here we discuss a class of models which are lacking such a direct relation,
but whose \bc s can nevertheless be classified completely.
The simplest example of this class is provided by the (critical) \tsp, 
for which the \chira\ \chir\ is the so-called $\chirw_3$-algebra. 
Potts models play a prominent role in the study of order-disorder transitions
and of high/low temperature duality, they are used for studying critical
percolation and linear resistance networks \cite{foka,card12}, and they are
of experimental interest e.g.\ for describing the
adsorption of monolayers on substrates \cite{beop}. Therefore
(as well as for the sake of definiteness), we first focus our attention to this
model, deferring a discussion of other theories to the end of this letter.

As has been established in \cite{fuSc6}, the classification of 
the conformally invariant \bc s for an arbitrary \cft\ naturally proceeds in
three steps. First one lists all \auto s of the fusion rules \caN\
which preserve the conformal weights $\Delta_\lambda$ (not only modulo 
integers, as would e.g.\ be required in the case of
the torus partition function). The second step consists in implementing such
\auto s on the spaces of \cb s. And finally one has to find all solutions for
certain scalar factors (which possess an interpretation as reflection
coefficients for bulk fields on the disk) that are compatible with the
factorization constraints. In string theory terms, this last step amounts
to identifying the possible types of Chan\hy Paton charges.
(In string theory, each such Chan\hy Paton charge
comes with its own multiplicity. E.g.\ in the uncompactified ten-dimensional
type I superstring theory, there is a single Chan\hy Paton charge 
with multiplicity 32.)

Concerning the first step, we recall that --
labelling the primary fields by $\lambda$ and assigning the special label
$\vac$ to the vacuum field -- a fusion rule \auto\ $\om$ satisfies 
$\N{\om(\lambda)}{\om(\mu)}{\,\ \ \ \om(\nu)}\eq\N\lambda\mu\nu$
and $\om(\vac)\eq\vac$. Using the Verlinde relation between the fusion
coefficients $\N\lambda\mu\nu$ and the modular \smat\ $S$ of the theory,
$\om$ must in particular preserve the \q \dim s $d_\lambda\eq\SD\lambda\vac/
\SD\vac\vac$. In the \tsp\ we employ a special
notation for the primaries, which together with the \q \dim s and
conformal weights is presented in the following table.
  \be \begin{tabular}{lccr} \hline \hline \Nxl22
  \multicolumn1c {$\lambda$} & Kac labels & $d_\lambda$ &
  \multicolumn1c {$\Delta_\lambda$} \\[.13em]
  \hline \Nxl20
  $\vac$   & $(11)\oplus(41)$ &  $1$  &  $0$       \\[.2em]
  $\one$   & $(43)$           &  $1$  &  $2/3$     \\[.2em]
  $\two$   & $(43)$           &  $1$  &  $2/3$     \\[.2em]
  $\three$ & $(21)\oplus(31)$ & $\frac12\,(1{+}\sqrt5)$ &  $2/5$  \\[.2em]
  $\four$  & $(33)$           & $\frac12\,(1{+}\sqrt5)$ &  $1/15$ \\[.2em]
  $\five$  & $(33)$           & $\frac12\,(1{+}\sqrt5)$ &  $1/15$ \\[.2em]
  \hline\hline \end{tabular}\ee
Thus from the \q \dim s $d_\lambda$ we already learn that $\om$ has to permute 
the elements of $\{\one,\two\}$ and of $\{\three,\four,\five\}$. 
(As is also already apparent from the values of $d_\lambda$, the full modular
\smat\ is just the tensor product of the \smat\ for the $\zet_3$ fusion rules
and the one for the Lee\hy Yang fusion rules \cite{card9}.)
Inspection of the specific fusion rules
  \be  \one\star\one=\two\,, \qquad  \one\star\two=\vac\,,\qquad
  \one\star\three=\four\,, \qquad  \one\star\four=\five  \ee
and
  \be  \three\star\three = \vac + \three \,, \qquad
  \four\star\four = \two + \five \,, \qquad
  \five\star\five = \one + \four  \ee
then tells us that \caN\ has precisely two \auto s: either $\om\eq\id$
is the identity, or else $\om\eq\cc$ acts as
  \be  \cc:\quad  \vac\mapsto\vac\,,\quad \one\leftrightarrow\two\,,\quad
  \three\mapsto\three\,,\quad \four\leftrightarrow\five \,,  \ee
which is just charge conjugation. Both \auto s 
indeed preserve the conformal weights, $\Delta_{\om(\lambda)}\eq\Delta_\lambda$.

The next task is to implement the fusion rule \auto\ $\om$ on all 
\cb s $V$ of the \cft\ on arbitrary surfaces. This means that 
for arbitrary choices of $\lambda_1,...\,,\lambda_m$ and
for every value of the insertion points and of the moduli of
the surface we must provide a family of associated isomorphisms
$\Theta_\om^{(\vec\lambda)}$ between the vector bundles
$V_{\lambda_1 \ldots \lambda_m}$ and $V_{\om(\lambda_1)\ldots\om(\lambda_m)}$
of blocks. 
Up to the restrictions originating from factorization, for the specification
of such maps it is sufficient to construct an implementation of $\om$ on all
the modules (\rep\ spaces) \hl\ 
of the \chira\ \chir. That is, we only need to prescribe
a family of maps $\oml{:}\;\hl\,{\to}\,\hlo$ that is consistent with the \cb\ 
structure. (Via state-field correspondence, this also induces a map on the 
chiral vertex operators, compare also \cite{reSC}.)

We first consider the vacuum sector \hvac; as a Virasoro module it
decomposes into irreducible submodules
as $\hvac\eq\hvaca\,{\oplus}\,\hwa$, where $\hvaca\,{\equiv}\,\hvacA$ is the
vacuum sector of the \teim\ -- i.e., in terms of the A-D-E classification of
minimal models, of the A-type model at the same value
$c\eq4/5$ of the conformal central charge -- while \hwa\ is the \vir-family 
that provides the spin-3 current of $\chirw_3$.
Preservation of the \vira\ means that $\omvac$ acts as the identity on
\hvaca, and then the $\chirw_3$ commutation relations 
(or equivalently, the fusion rule $\ja\,{\star}\,\ja\eq\vaca$ of the \teim)
imply that $\omvac$ acts as $\pm\,\id$ when restricted to \hwa.
Moreover, inspecting the action of \chir\ on the non-selfconjugate sectors
one sees that the two sign choices correspond precisely to $\om\eq\id$ and
$\om\eq\cc$, \resp. We conclude that each of the two fusion rule \auto s 
possesses a unique implementation $\omvac$ on \hvac.

Next we show that the same result applies to the implementations $\oml$ on all 
other sectors \hl\ as well. We first determine the implementation only up to
a non-zero over-all scalar factor, which takes into account the fact that
the \hws\ of \hl\ is unique only up 
to a scalar multiple. Now the \chir-modules \hl\ other than \hvac\ and
$\hil_\three$
all consist of a single Virasoro module. {}From the invariance of the
\vira\ it therefore follows that in these cases $\oml$ is already defined by
its action on the \hws\ of \hl.
Furthermore, the image of the \hws\ of \hl\ under $\oml$ must be the \hws\ of
\hlo, and hence $\oml$ is uniquely determined (up to the freedom of
changing the \hws s by non-zero scalar factors). In short, 
for $\lambda\iN\{\one,\two,\four,\five\}$ these maps $\oml$
just `exchange' the whole modules \hl\ and \hlo.
As for the remaining sector $\hil_\three$ which decomposes into Virasoro 
modules as $\hil_\three\eq\hwb\,{\oplus}\,\hwc$, we note that the \hwv s
of the two components have distinct conformal weights ($2/5$ and $7/5$,
\resp). Invariance of the \vira\ therefore implies that $\om_\three$ acts
as a multiple of the identity when restricted to \hwb\ or \hwc. 
Up to an over-all normalization, we can set $\om_\three\eq\id$ on \hwb;
using also the transformation property of the rest of the \chira,
we then see that we must have $\om_\three\eq{\pm}\,\id$ when restricted to \hwc,
and that the choice of sign again precisely corresponds to the alternative of 
having $\om\eq\id$ and $\om\eq\cc$, \resp. 

This finishes the determination of the `\auto\ type' \cite{fuSc6} of \bc s. 
To complete the classification, we finally need to determine, for each of the 
two \auto\ types separately, the possible values of scalar 
factors, i.e.\ of Chan\hy Paton charges. These charges
may be regarded as reflection coefficients $\rc A\lambda\vac$ for bulk fields 
on the disk \wrt the vacuum boundary field \cite{prss3,fuSc5}; they must 
respect the factorization constraints that implement \cite{lewe3,prss3} the 
compatibility of a \cft\ on surfaces of different topology. 
Let us from now on take the torus partition function in the bulk to be
given by charge conjugation (when one deals 
instead with the diagonal torus amplitude, the results for $\om\eq\id$
and $\om\eq\cc$ are interchanged). Then the \bc s with $\om\eq\cc$ are 
governed \cite{card9} by the fusion rule \alg\ \caN. In particular, they
may be labelled as $A\eq(\cc;\lambda)$ by the primary fields, so there are
six different \bc s, and the reflection coefficients are given by the
\onedim\ \irrep s of \caN, i.e.\ by the quotients
  \be  \rcc\lambda\mu\vac = \SD\lambda\mu / \SD\lambda\vac  \ee
of \smat\ elements. For $\om\eq\id$ the situation is less familiar. But again 
the \bc s are governed by some \findim\ commutative \alg\ \caD\ \cite{fuSc6}. 
To construct \caD, we first observe that the fusion rules of the \tsp\ can be 
obtained from those of the \teim\ by combining fields that are related by 
fusion with the field $\pja$. More precisely,
  \be  \N\lambda\mu\nu = \NI\lambda\mu\nu + \NJ\lambda\mu\nu   \labl n
for $\ja\,{\star}\,\dot\nu\nE\dot\nu$, while in the case of $\ja\,{\star}\,
\dot\nu\eq\dot\nu$, i.e.\ $\nu\iN\{\one,\two,\four,\five\}$, the prescription is
more involved (see \cite{scya6}); here on the \rhs\ $\lambdad\eq\lambda$ when
\hl\ is \vir-irreducible, while otherwise $\lambdad$ stands for the label 
of any of the two \vir-irreducible subspaces of \hl.
Now when expressed in terms of the fields of the \teim, the action of 
the implementing map $\cc_\vac$ amounts in particular to mapping $\pja$
to $-\pja$. It follows that the structure constants of the \alg\ \caD\ 
are given by a formula analogous to \erf n:
  \be  \ND\lambda\mu\nu = \NI\lambda\mu\nu - \NJ\lambda\mu\nu \,.  \labl-
Inspecting the fusion rules $\NI\lambda\mu\nu$, this relation can be seen to
imply that \caD\ is isomorphic to the fusion rule \alg\ of 
the Lee\hy Yang non-unitary minimal model. In particular,
\caD\ is \twodim; its natural basis corresponds to the fields $\phi_\vac$ and
$\phi_\three$, while its two \onedim\ \irrep s are naturally labelled by the
$\zet_2$-orbits $\{(22),(32)\}$ and $\{(44),(42)\}$ 
 -- to which we will refer as $\zwei$ and $\vier$, \resp\ --
of sectors of the \teim\ that do not appear in the \tsp.
(Note that according to \erf- the sign of $\ND\lambda\mu\nu$ depends on 
the choice of representatives $\lambdad$ of $\lambda$; this does not constitute,
however, any physical ambiguity, because these choices must be matched by 
analogous sign choices for the relevant Ishibashi \cite{card9} states,
in such a way that the annulus partition function is non-negative.)
It follows in particular that
  \be  \rci\lambda\mu = \SI\lambda\mu / \SIv\lambda   \ee
with $\lambda\iN\{\zwei,\vier\}$ and $\mu\iN\{\vac,\three\}$. 
Explicitly, we have
  \be \bearl  \rci{\zwei}\vac = 1 = \rci{\vier}\vac \,, \nxl5 
  \rci{\vier}\three = \frac12\,(1{+}\sqrt5) = -1/\rci{\zwei}\three \,. \eear \ee

We can summarize: For the \tsp\ there are eight distinct conformally
invariant \bc s $A\eq(\om;\lambda)$, six of them with $\om\eq\cc$ and two with 
$\om\eq\id$. It is then straightforward to compute
the partition function on an annulus of modular parameter $t$
with \bc s $A$ and $B$ and expand it as
  \be  Z_{AB}^{}(t) = \sum_\lambda Z_{AB}^\lambda\,\chii_\lambda^{}(\ii t/2)
  \labl z
\wrtt characters $\chii_\mu$.
Inspection then allows for the following identification \cite{card9,afos} 
with \bc s in the spin chain formulation \cite{vorr}.
For $\om\eq\cc$ the conditions labelled by elements of the 
$\zet_3$-orbit $\{\vac,\one,\two\}$ are the three possible {\em fixed\/} \bc s, 
while the other $\zet_3$-orbit $\{\three,\four,\five\}$ corresponds to 
{\em mixed\/} \bc s where two out of the three possible values of the spin
variable are allowed; $\om\eq\id$ with $\lambda\eq\vier$ yields {\em free\/}
\bc s. Finally, the \bc\ $\om\eq\id$ with $\lambda\eq\zwei$, which was
recently discovered in \cite{afos}, does not possess an obvious interpretation
in terms of the spin variable of the lattice model.

It is not difficult to verify that the \bc s obtained above obey the
usual consistency conditions. First, as already observed in \cite{card9,afos},
the numbers $Z_{AB}^\mu$ defined by \erf z are non-negative integers.
Furthermore, they obey the associativity relation
  \be  \sum_\mu Z^\mu_{AB}\, Z^{\mu^+_{}}_{CD}
  = \sum_\mu Z^\mu_{AC^+_\P}\, Z^{\mu^+_{}}_{B^+_\P D} \,. \labl5
And finally, as matrices in the labels $A,B$ they satisfy
  \be  A^\mu\, A^\nu= \sum_\lambda \Ntot\mu\nu\lambda\, A^\lambda \,; \labl6
the number
$\Ntot\mu\nu\lambda$ equals the fusion coefficient of the \tsp\ when
all the labels $\mu,\nu,\lambda$ refer to sectors that are present in the 
model, while it is a linear
combination of fusion coefficients of the \teim\ when precisely two out of
these labels refer to the $\zet_2$-orbits $\zwei$ or $\vier$, and is zero
otherwise.

Let us now look at these results for the Potts model from the perspective of 
general \cft. As a matter of fact, the situation encountered above is but a 
special
case of the following general setup. One deals with a `D-type' model which is
obtained from an associated `A-type' model -- another \cft\ with the same
value of $c$ -- by extending the \chira\ by a primary field
$\pJ$ that has the simple (A-model) fusion rules $\J\,{\star}\,\J\eq\vacA$.
(Such a field is called \cite{scya6} an integer spin simple current;  
in the Potts case above it is the spin-3 current of $\chirw_3$.) Thus the 
vacuum sector of the D-model decomposes as $\hvac\eq\hvacA\,{\oplus}\,\hJ$
into irreducible modules of the non-extended \chira.
Moreover, one can distinguish two types of sectors \hl: those which are
reducible as modules over the \chira\ of the A-model and those which are
irreducible; we refer to them as A-reducible and A-irreducible sectors, \resp.
For A-reducible sectors one has in fact $\hl\eq\hil_\lambdad\,{\oplus}\,
\hil_{\J\star\lambdad}$ with $\J\,{\star}\,\lambdad\nE\lambdad$, while
A-irreducible sectors appear only for $\J\,{\star}\,\lambdad\eq\lambdad$ and 
always come in pairs, so we denote them by $\hL^\pm$ (the sectors
$\hL^+$ and $\hL^-$ are isomorphic as \rep\ spaces of
the \chira\ for the A-model, but not of the one for the D-model).

It is also known \cite{fusS6} that in such D-models there is always a 
non-trivial fusion rule \auto\ $\Om$ preserving conformal weights. 
On A-reducible sectors, in particular on the vacuum sector, this map $\Om$
can be implemented uniquely (up to over-all scalar factors) as
  \be  \Oml = \id^{}_{{\cal H}_\lambdad} \oplus -\id^{}_{{\cal H}_{\J\star
  \lambdad}} \,,  \labl1
while the implementations on A-irreducible sectors read
  \be  \Oml{}^\pm:\quad \hL^\pm \to \hL^\mp \,,  \labl2
with \hws s mapped to \hws s. Often 
$\Om$ will be the only non-trivial implementable fusion rule
\auto\ (and thus coincide with charge conjugation when $\cc$ is non-trivial).
In the sequel we restrict our attention to models where this is the case.

We denote the fusion rule \auto\ that describes the torus partition function
by $\pi$. As in any \cft, for $\om\eq\pi$ the
\bc s are governed by the fusion rule \alg\ \caN\ of the D-model. We find that
for $\om\eq\pi\,{\circ}\,\Om$ the role of \caN\ is taken
over by a commutative \alg\ \caD\ with the following properties. The dimension 
of \caD\ equals the number \NO\ of A-reducible sectors, and there is a basis
of \caD\ labelled by those sectors; in this basis the structure constants read
  \be  \ND\lambda\mu\nu = \NA\lambda\mu\nu - \NAJ\lambda\mu\nu \,.  \labl3
Finally, the inequivalent \onedim\ \irrep s of \caD\ are labelled by the
(\NE\ many) $\zet_2$-orbits $[\lambda]\eq\{\lambdad,\J{\star}\lambdad\}$
of the A-model that do not appear in the spectrum of the D-model; they are
given by
  \be  \rcA\lambda\mu = \SA\lambda\mu / \SAV\lambda \,.  \ee
We can also show that \caD\ is associative and (with an appropriate
correlated choice of the signs in \erf3 for conjugate sectors) has a
conjugation such that $\ND\lambda\mu\vac\eq\delta_{\lambda,\mu^+}$.
This implies that \caD\ is semisimple and hence all its \irrep s are \onedim. 
It follows e.g.\ that the number
of inequivalent \irrep s of \caD\ equals its dimension, i.e.\
  \be  \NE = \NO \,.  \labl N

Using identities among \smat\ elements such as those employed in \cite{fuSc5},
by direct computation one can again check that these \bc s satisfy the
usual consistency relations. That is, the annulus coefficients $Z_{AB}^\mu$ 
defined as in \erf z are non-negative integers and they satisfy the relations 
\erf5 and \erf6, with the coefficients $\Ntot\mu\nu\lambda$ in \erf6 of a form 
completely analogous to the case of the \tsp.

Besides the \tsp, other examples of this structure
are given by the D$_{\rm even}$-type $\liefont{su}(2)$ \wzwts\
and by free bosons; in the latter case the A-model
is the corresponding $\zet_2$-orbifold. As an illustration,
we note that the distinction between $\om\eq\cc$ and $\om\eq\id$ is
a direct generalization of the situation in the \cft\ of a free boson $X$.
In that case the role of \hwa\ is taken over by the $\liefont u(1)$-current
$j\eq\ii\partial X$, so that in particular $\om\eq\cc$ which changes the sign of
$j$ corresponds to Neumann, while $\om\eq\id$ corresponds to Dirichlet \bc s.
(This applies
for the case of the charge conjugation torus partition function $\pi\eq\cc$, 
while for the diagonal torus partition function
$\pi\eq\id$ it is the other way round.) 

Similar constructions for the case of several
free bosons $X^i$ classify the \bc s that correspond to D-branes in string
theory. It is natural to use the same nomenclature,
i.e.\ `Neumann \auto\ type' for $\om\eq\pi$ and `Dirichlet \auto\ type'
for $\om\eq\pi\,{\circ}\,\Om$, also for any other model of the type described 
above. Thus in particular in the \tsp\ the fixed and mixed
\bc s are analogues of Neumann conditions, while the free
\bc\ and the new one of \cite{afos} are analogues of Dirichlet conditions.
Note that $T$-duality amounts to exchanging the meaning of 
Neumann and Dirichlet conditions.  We see that $T$-duality
does not act one-to-one between \bc s, but rather between suitable
orbits of them; e.g.\ in the \tsp\ it maps fixed to free conditions and mixed
conditions to the new one of \cite{afos}, and vice versa. 
We expect that these orbits come from Galois transformations \cite{fgss}
of S-matrix elements.

When the free boson $X$ is compactified on a circle of rational radius squared,
then the Dirichlet-type \alg\ \caD\ is \twodim\ and is isomorphic to the
$\zet_2$ fusion rules, while for the non-diagonal $\liefont{su}(2)$ \wzwt\ at
level $4\ell$ the \alg\ \caD\ has dimension $\ell$ and turns out to be
isomorphic to the fusion rule \alg\ of the 
non-unitary minimal model of type $(2\ell{+}1,2)$.

Another class of such D-models is given by the unitary
minimal models of conformal
central charge $c\eq1{-}6/(m{+}1)(m{+}2)$ with $m\eq4\ell$
for some $\ell\iN\zet_{>0}$, with the \chira\ extended by the field 
$\pJ\,{\equiv}\,\phi_{(m1)}$ of conformal weight $\Delta_\J\eq m(m{-}1)/4$.
(The similar series with $m\eq4\ell{+}1$ can be treated analogously.)
These have a total of $2\ell(\ell{+}2)$ sectors, among them
$2\ell^2$ A-reducible modules and $2\ell$ pairs of A-irreducible ones.
Inspecting the fusion rules of these models, we find
that just like in the \tsp\ (which is obtained for $\ell\eq1$), for any
$\ell$ there are
only two possible fusion rule \auto s, namely the identity and charge
conjugation. Both of them preserve conformal weights, and up to scalar factors 
they can be implemented uniquely on the modules \hl, in the way described in
\erf1 and \erf2.

\newpage
 \newcommand\wb{\,\linebreak[0]} \def\wB {$\,$\wb}
 \newcommand\Bi[1]    {\bibitem{#1}}
 \def\ijmp  {Int.\wb J.\wb Mod.\wb Phys.\ A}
 \def\jopa  {J.\wb Phys.\ A}
 \def\nupb  {Nucl.\wb Phys.\ B}
 \def\phlb  {Phys.\wb Lett.\ B}
 \def\phrb  {Phys.\wb Rev.\ B}
\def\jf      {J.\ Fuchs}
\def\modinv  {modular invarian}
  \renewcommand\J[5]   { {\sl #5}, {#1} {#2} ({#3}) {#4} }
  \newcommand\Prep[2]  {{\sl #2}, preprint {#1}}
  
\end{document}